# THE IMPACT OF DATA REPLICATION ON JOB SCHEDULING PERFORMANCE IN HIERARCHICAL DATA GRID

Somayeh Abdi[1], Hossein Pedram[2], Somayeh Mohamadi[3]

[1]Department of Computer Engineering, Science and research branch Islamic Azad University, Tehran, Iran
s.abdi@srbiau.ic.ir

[2] Computer Engineering Departments, Amirkabir University, Tehran, Iran
pedram@aut.ic.ir

[1]Department of Computer Engineering, Ghasre Shirin branch Islamic Azad University, Ghasre Shirin, Kermanshah, Iran
s.mohamadi@srbiau.ic.ir

## ABSTRACT

*In data-intensive applications data transfer is a primary cause of job execution delay. Data access time depends on bandwidth. The major bottleneck to supporting fast data access in Grids is the high latencies of Wide Area Networks and Internet. Effective scheduling can reduce the amount of data transferred across the internet by dispatching a job to where the needed data are present. Another solution is to use a data replication mechanism. Objective of dynamic replica strategies is reducing file access time which leads to reducing job runtime. In this paper we develop a job scheduling policy and a dynamic data replication strategy, called HRS (Hierarchical Replication Strategy), to improve the data access efficiencies. We study our approach and evaluate it through simulation. The results show that our algorithm has improved 12% over the current strategies.*

## 1. INTRODUCTION

In the increasing demand of scientific and large-scale business application, a large amount of data are generated and spread for using by users around the world. Many good examples can be listed such as High Energy Physics, meteorology, computational genomics which processes and results large amount of data. Such data cannot be stored centralized in any site but distributed among centre around the world [1]. Data grid tries to store this data in decentralize sites and then for each application retrieves it from these sites. Data grid as an important branch of grid computing focuses on supporting an efficient management mechanism for controlled sharing and large amounts of distributed data. In Data Grid, for each incoming job, the Grid scheduler decides where to run the job based on the job requirements and the system status. Scheduling jobs to suitable grid sites is necessary because data movement between different grid sites is time consuming. If a job is scheduled to a site where the required data are present, the job can process data in this site without any transmission delay for getting data from a remote site.

Data replication is an important optimization step to manage large data by replicating data in geographically distributed data stores. When users' jobs access a large amount of data from remote sites, dynamic replica optimizer running in the site tries to store replicas on local storage for future possible repeated requests.

We proposed a new job scheduling algorithm and data replication policy that reduce job execution time by reducing job data access time. Our new scheduling policy considers the locations of required data and the job queue length of a computing node. We develop a replication strategy, called HRS (Hierarchical Replication Strategy). It takes into account bandwidth as an important factor for replica selection and replica placement. It also increases the chances of accessing data at a nearby node.

The rest of the paper is organized as follows, in section 2 we present a summary of existing and related work. In section 3, a hierarchical structure is proposed for data grid based on classification of networks, along with a scheduling algorithm and replication algorithm for this structure, Section 4 describes our experiments and the results achieved followed by conclusion in section 5.

## 2. RELATED WORK

There are some recent works that address the problem of scheduling and/ or replication in Data Grid as well as the combination between them.
In [3], it considers two centralized and decentralized replication algorithms. In centralized method, replica master uses a table that ranks each file access in descending order. If a file access is less than the average, it will be removed from the table. Then it pop files from top and replicates using a response-time oriented replica placement algorithm. In the decentralized method, every site records file access in its table and exchange this table with neighbours. Since every domain knows average number of access for each file and then deletes those files whose access is less than the average, and replicates other files in its local storage.
In [5], an algorithm for a two-level hierarchical structure based on internet hierarchy (BHR) has been introduced which only considers dynamic replication and does not consider scheduling. Nodes in the first level are connected to each other with high speed networks and in the second level via internet. The algorithm replicates the file to the site if there is enough space. Next it, accesses the file remotely if the file is available in the sites that are in the same region. Otherwise it tries to make available space by deleting files using LRU (Least Recently Used) method, and replicates the file. It assumes that master site always has a safe copy of file before deleting.
In [6], a structure with few networks connected via internet has been presented and an algorithm similar to [5], along with scheduling is proposed. For replicating a file, first computes the total transfer time, then it selects the best node with shortest transfer time.
In [7] authors introduce dynamic replication placement (RP) that categorizes the data based on their property. This category is used for job scheduling and replication. Then a job is allocated to a site which has the file in the required category, this leads to reduce the cost for file transfer.
In [8] a Genetic Algorithm based co-scheduling of data and jobs of independent nature was proposed; the GA is executed to converge to a schedule by looking at the jobs in the scheduler queue as well as the replicated data objects at once. A performance overhead might be incurred by the system as a result of delaying the replication of the data until the scheduling time. The authors also use an objective function that assumes infinite availability of storage for all data objects which is infeasible in a realistic grid setting.

## 3. THE PROPOSED METHOD

In this section, we present hierarchical network structure and then we proposed two strategies for job scheduling and data replication by considering the hierarchical network structure.

### 3.1. NETWORK STRUCTURE

The proposed Data Grid structure is shown in Figure 1. This structure has two levels. Regions comprise the first level and sites that are located in a region comprise the second level. A region represents an organization unit which is a group of sites that are geographically close to each other, each region comprises the computers which are connected by a high bandwidth. We define two kinds of communications between sites, inter-communication and intra-communication. Intra-communication is the communication between sites within the same region and inter-communication is the communication between sites across regions. Network bandwidth between sites within a region will be larger than across regions.

In communication networks, the performance of system is underlying available network bandwidth and data access latency, especially in networks that hierarchy of bandwidth appears. Therefore, to reduce access latency and to avoid WAN bandwidth bottleneck, it is important to reduce the number of inter-communications.

Data Grid Information Service (DGIS) providing resource registration services and keeping track of a list of resources available in the Data Grid. The Grid Scheduler can query this for resource contact, configuration, and status information; Resource discovery identify resources that can be used with their capability through Data Grid Information Service.

In this structure we apply distributed dynamic data replication infrastructure. Replica Manager at each site manages the data movement between sites. When a job is assigned to Local Scheduler of site, those required data that do not exist in local site will be transfer to it. Data request by the job is generated as soon as job is scheduled into Local Scheduler queue and Replica Manager controls data movement at each site.

In proposed structure there is a centralized Replica Catalogue. It is responsible for indexing available files on the resources and handles queries from users and resources about location of replicas.

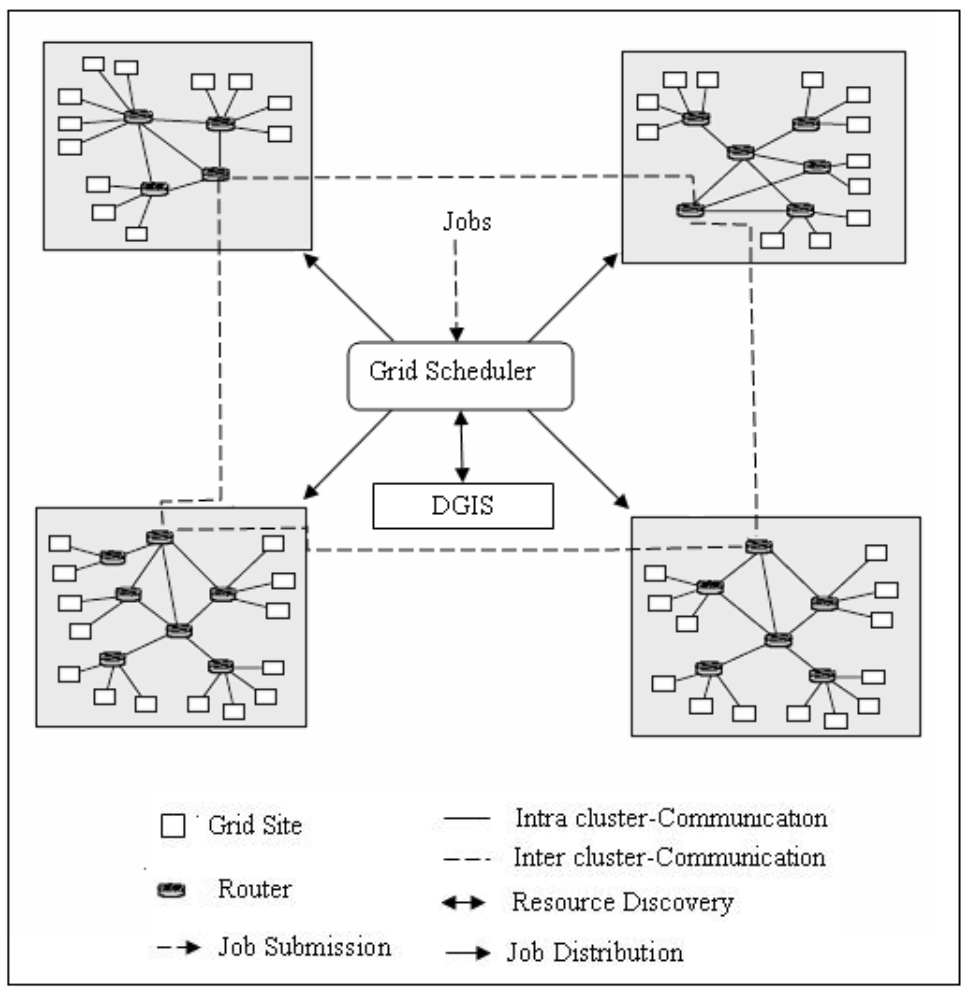

Figure 1. Data Grid Architecture.

## 3.2. SCHEDULING ALGORITHM

Our new scheduling policy considers the locations of required data, computing capacity and number of jobs that allocate on computing nodes. For efficient scheduling of any job, the algorithm determines the best site and then submit job to Local scheduler. A best site is a site that holds most of the requested files (from size point of view). This will significantly reduce total transfer time, and reduce the job execution time.

Each job requires some replicas and the replicas is needed to execute job j are represented as: $R_j$ = {$LFN_1$, $LFN_2$,…., $LFN_n$}. We assume that $S_j$ is total size of the requested files available in site j, $C_i$ is computing capacity (in MIPS[1]) of site i and $RelativeLoad_i$ is relative load of site i.

---

[1] . Million Instruction Per Second

$$S_S = \sum_{\text{for all available } LFN_i \text{ in site S}} |LFN_i| \quad (1)$$

$$\text{RelativeLoad}_i = \text{SizeofJobs}_i / C_i \quad (2)$$

SizeofJobs$_i$ is the lengths of queued jobs (in MIPS) on site i and $|LFN_i|$ present size of file LFN$_i$.

The scheduling algorithm can be summarized as follow:

1- Compute $S_C$ for each site from (1)

2- Select the best site $S_{\max} = \underset{i=1}{\overset{q}{MAX}} \, S_i$; where q is the number of site (i.e. site with largest available requested data files)

3- If there are several sites with most available data, select the site with minimum relative load from (2).

When a job is assigned to local scheduler, the Replica Manager transfers all the requested files that are not exist in local site. The objective of file replicating is to transferring required data to local site before job execution. Therefore, data replication improves job scheduling performance by reducing job execution time.

The proposed scheduling algorithm takes into consideration the requested data files. Additionally, where there are the same data requirements in several sites (from size point of view); it takes into consideration the computing capacity of sites with a view to reducing the queue waiting time and relative load of sites.

### 3.3. HIERARCHICAL REPLICATION STRATEGY

After a job is scheduled to site Sj , the requested data will be transferred to Sj to become replicas. HRS (hierarchical replication Strategy) determines wich replica will be transferred to Sj and how to handle this new replica. HRS considers the bandwidth between the regions as the main factor for replica selection / deletion.
When each site store a new replica, replica manager send a file register request to RC and then RC add this site to the list of sites that holds the replica. Replica Manager controls data transferring in each site and provide a mechanism for accessing the Replica Catalogue. Data Grid Architecture for Replica Catalogue and Replica Manager is shwon in in Figure 2.

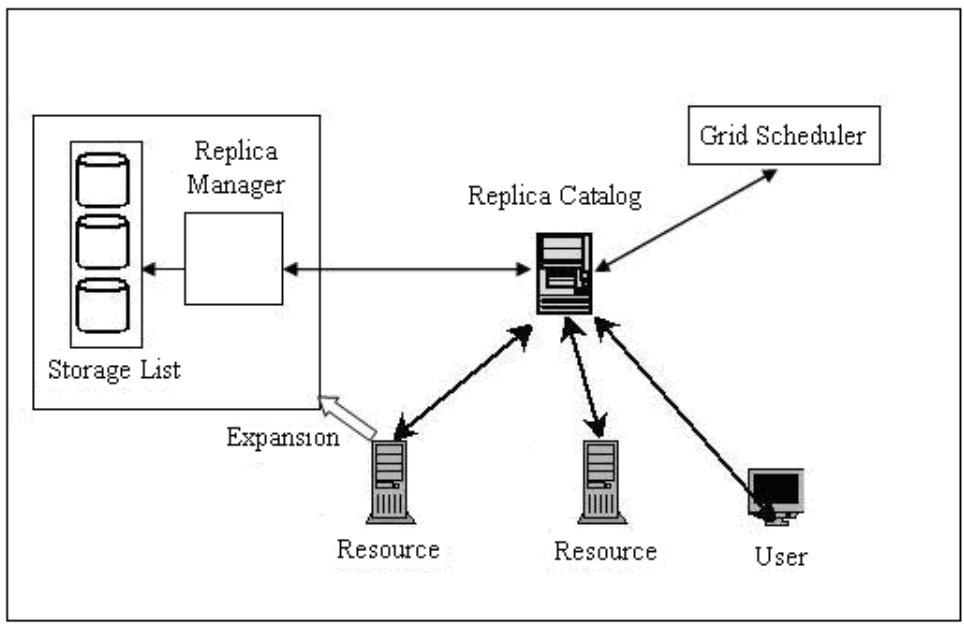

Figure 2.  Data Grid Architecture for replica cataloge and replica manager.

For each required file of the job, Replica Manager controls the existence of the file in local site. If file doesn't exist in the local site HRS first search the file in local region. If the file duplicated in the same region, then it will create a list of candidate replicas and selects a replica with the maximum bandwidth available for transferring it. If there is enough space for new replica, then store it in the local site, otherwise it is only stored in the temporary buffer and will be deleted after the job completes.

If the file doesn't exit in the same region, then HRS create a list of replicas in other regions and select the replica with the maximum bandwidth available for transferring it. If there is enough space for new replica, then store it in the  local site, otherwise occupided space will be released to make engough room for this new replica. First, it removes the replicas that already exist in other sites in the same region based on LRU(least recently used) replacement algorithm. After all these replicas are deleted, if the space is still insufficient, the HRS use LRU replacement algorithm to  delete replica in local  storage which is duplicated in other regions , till it has enough room for the new replica.

## 4.  SIMULATIONS

Gridsim is used as the simulation tool to evaluate the performance of the proposed replication and scheduling algorithms. The Java-based GridSim discrete event simulation toolkit provides Java classes that represent entities essential for application, resource modelling, scheduling of jobs to resources, and their execution along with management [9]. Its Java-based design makes it portable and available on all computational platforms. The components of Gridsim are as follow and also depicted in Figure 3.

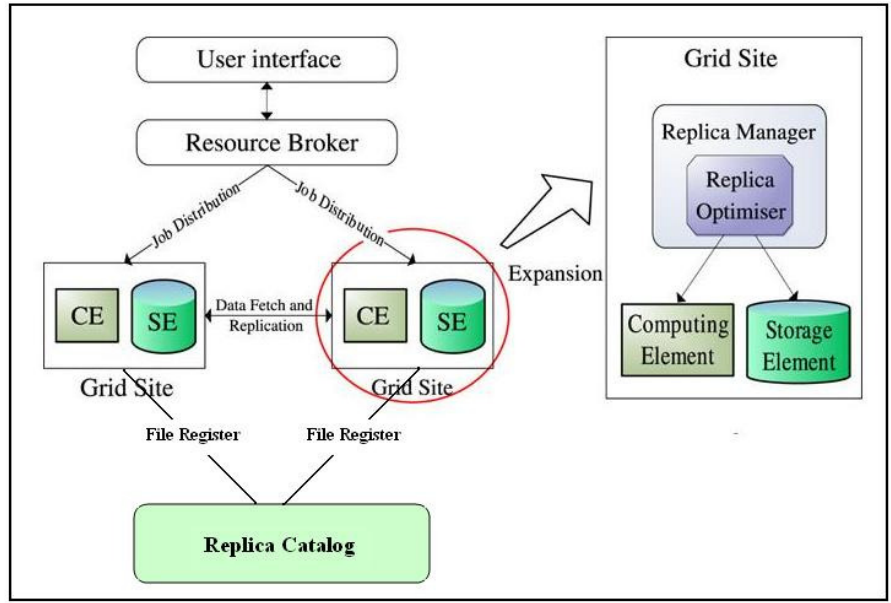

Figure 3. Simulator architecture

Resource Broker: It receives jobs from user and sends them to the best node according to proposed algorithm; Storage Element (SE): Storage resource in grid; Computing Element (CE): Computing resource in grid; Replica Manager:controls data transfering at each site; Replica Catalogue: stores list of sites that holds replicas.

Based on the scheduling algorithm the broker sends jobs to a node. Each job needs a list of files to run. Reducing file access time is the final objective of optimization algorithms.

### 4.1. Simulation Environment

There are four regions in our configuration and each region has an average of 13 sites, which all have CE with associated SE. Table1 specifies the simulation parameters used in our study. There are 5 job types; each job type requires 12 files to execute. While running, jobs were randomly picked from 5 job types, then submitted to the Resource Broker. Files are accessed sequentially within a job without any access pattern. To simplify the requirements, data replication approaches in Data Grid environments commonly assume that the data is read-only.

TABLE 1. Simulation parameters

| Topology Parameters | value |
|---|---|
| Number of region | 4 |
| Number of sites in each region | 13 |
| Storage space at each site | 10 GB |
| Connectivity bandwidth(WAN) | 10 Mbps |
| Connectivity bandwidth(LAN) | 1000Mbps |
| **Job parameters** | **value** |
| Number of jobs | 500 |
| Number of job types | 5 |
| Number of file accessed per job | 12 |
| Size of single file | 500 MB |
| Total size of files | 50GB |

### 4.2. Simulation results and discussion

HRS will be compared with LRU (Least Recently Used) and BHR (Bandwidth Hierarchy based Replication). The LRU algorithm always replicates and then deletes those files that have been used least recently. Figure 4 shows the Average job time based on changing number of jobs for 3 algorithms. Figure 5 shows the Average job time for 1000 jobs by 3 mentioned algorithms. HRS replication strategy uses the concept of "network locality" as BHR [5].

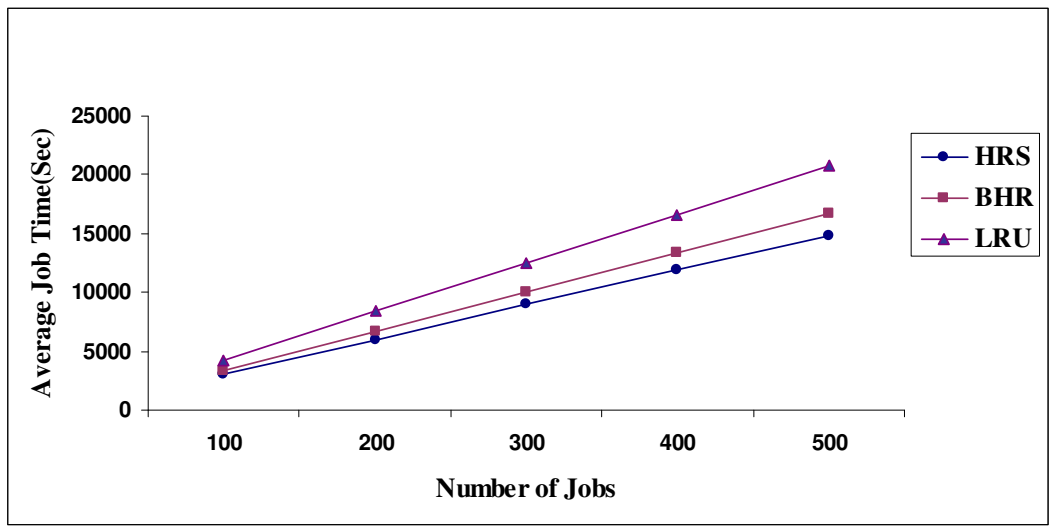

Figure 4. Average job Time based on varying number of jobs.

The difference between HRS and BHR is that required replica within the same region is always the top priority used in HRS, while BHR searches all sites to find the best replica and has no distinction between intra-region communication and inter-region comuuniation. It could be anticipated that HRS will avoid inter-region communications and be stable in hierarchical network architecture with variable bandwidth. Our method takes benefit from network level locality of BHR. Thus, total job execution time is about 12% faster using HRS optimizer than BHR.

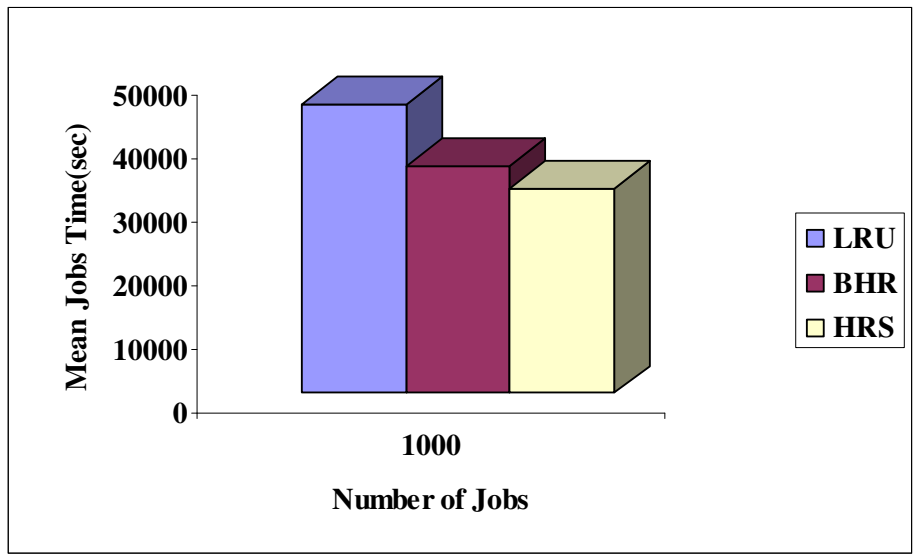

Figure 5. Average Job Time for 1000 jobs.

The job execution time is the Max{file transmission time, queue time} plus job processing time. Since the file transmission time is the most important factor to influence the job execution time for data-intensive jobs in data grids, the proposed scheduling algorithm with HRS can reduce the file transmission time effectively by virtue of valid scheduling and proper data replication, as can be seen from the experiments.

The average number of inter-communications for a job execution is illustrated in Figure 6. By selecting the best site based on location of required data by the job, the proposed scheduling algorithm with HRS can decrease the number of inter-communications effectively. Overall the simulation results with Gridsim show better performance (over 12%) comparing to current algorithms.

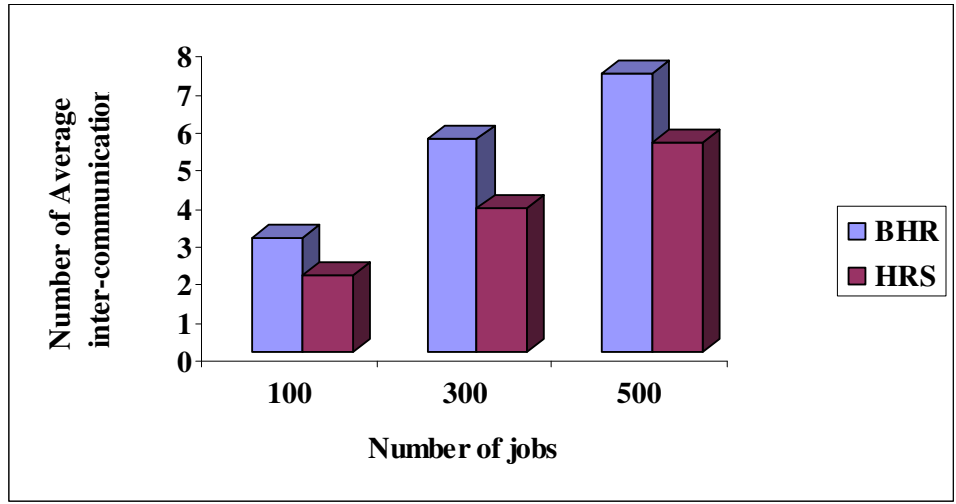

Figure 6. Average number of inter-communications

Figures 7 shows the average job time for 500 jobs. We compare HRS, BHR and LRU algorithms for varying inter-communication bandwidth. As inter-communication bandwidth increase 3 mentioned algorithms will converge. We can conclude that HRS strategy can be effectively utilized when hierarchy of bandwidth appears because BHR has no distinction between intra-region and inter-region. It could be anticipated that HRS will avoid inter-region communications and be stable in hierarchical network architecture with variable bandwidth.

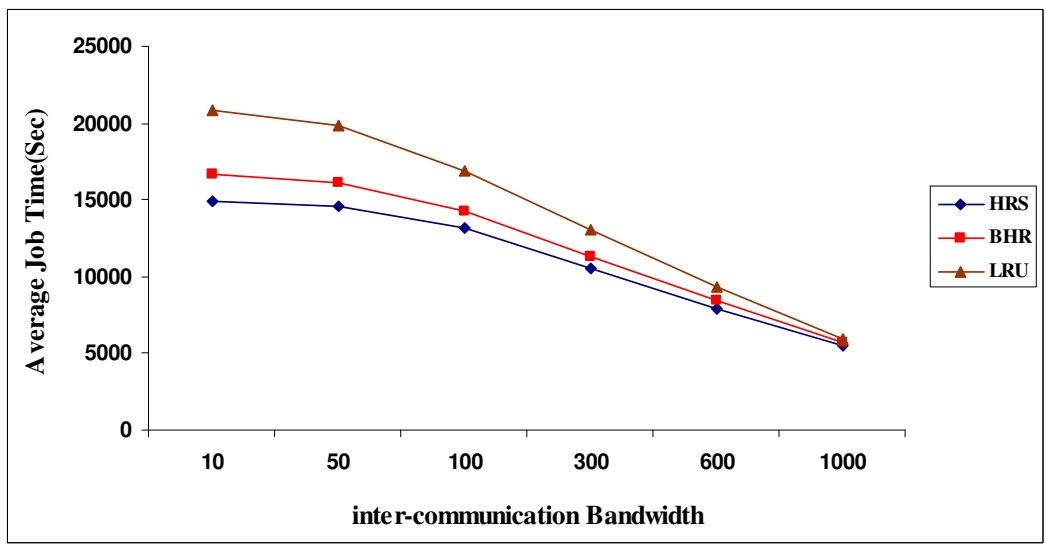

Figure 7. Average Jobs Time with varying inter-communication bandwidth for 500 jobs

Overall the simulation results with Gridsim show better performance (over 12%) comparing to current algorithms.

## 5. CONCLUSION AND FUTURE WORK

In this paper a hierarchical structure for dynamic replicating file and scheduling in data grids was proposed. To achieve good network bandwidth utilization and reduce data access time, we propose a job scheduling policy that considers computational capability, job type and data location in job placement decision. We study and evaluate the performance of various replica strategies. The simulation results show, first of all, that proposed scheduling algorithm and HRS both get better performances. Second, we can achieve particularly good performance with scheduling algorithm where jobs are always scheduled to site with most of the data needed, and a separate HRS process at each site for replication management. Experimental data show proposed scheduling algorithm with HRS replica strategy outperforms others combinations in total job execution time.